\documentclass[twocolumn,amssymb,fleqn]{revtex4} 
\usepackage{epsfig,amssymb,amsmath,graphicx,hyperref}  
\usepackage{color}

\usepackage{multirow} 
\usepackage{tabularx}

\usepackage{graphicx}
\graphicspath{{img/}}
\usepackage[caption=false]{subfig}
\newcommand{\be} {\begin{eqnarray}}
\newcommand{\ee} {\end{eqnarray} }

\begin{document}

\title{Intrinsic conformal order revealed in geometrically confined \\
long-range repulsive particles}
\author{Zhenwei Yao}
\email{zyao@sjtu.edu.cn}
\affiliation{School of Physics and Astronomy, and Institute of Natural
Sciences, Shanghai Jiao Tong University, Shanghai 200240, China}
\begin{abstract} 
Elucidating long-range interaction guided organization of matter is a
  fundamental question in physical systems covering multiple length scales.
  Here, based on the hexagonal disk model, we analyze the characteristic
  inhomogeneity created by long-range repulsions, and reveal the intrinsic
  conformal order in particle packings in mechanical equilibrium.  Specifically, we
  highlight the delicate angle-preserved bending of the lattice to match the
  inhomogeneity condition. The revealed conformal order is found to be protected
  by the surrounding topological defects. These results advance our
  understanding on long-range interacting systems, and open the promising
  possibilities of using long-range forces to create particle packings not
  accessible by short-range forces, which may have practical consequences.
\end{abstract}

\maketitle

\section{Introduction}

Long-range forces provide a unique mechanism to organize
elementary constituents into a host of structures of multiple length scales that are not accessible by short-range
forces~\cite{campa2014physics,levin2014nonequilibrium}, ranging from self-gravitating
systems~\cite{padmanabhan1990statistical, benetti2014nonequilibrium},
plasmas~\cite{thomas1994plasma,Levin2002},
elastic~\cite{Landau1986, audoly2010elasticity} and hydrodynamic
systems~\cite{chattopadhyay2009effect, dallaston2018discrete}, to the exceedingly rich electrostatic
phenomena in electrolyte solutions~\cite{Holm2001, Walker2011, lee2015direct, shen2019universal}. Despite their ubiquity, the long-range coupling among the
constituents imposes grand challenge for gaining insights into the resulting
intriguing physics~\cite{schweigert1995spectral, ouguz2011helicity,
ribeiro2014ergodicity}. Studies on the packings of geometrically confined
particles reveal the strong connection between long-range repulsion and
inhomogeneity~\cite{Berezin1985,yao2013topological,soni2018emergent,silva2020formation,PhysRevE.108.025001}.
Analyzing the inhomogeneity phenomenon thus offers a promising perspective to
study the fundamental physics of long-range interactions.  This approach has
proven fruitful, yielding the important concepts of curvature~\cite{Mughal2009,
yao2013topological,soni2018emergent}, topological
defects~\cite{Lai1999,nelson2002defects, Mughal2007,Klumov2022}, and conformal
crystal~\cite{pieranski1989gravity,rothen1993conformal,rothen1996mechanical,wojciechowski1996minimum}
to uncover the featured geometric structures.

Previous studies show that long-range interactions create inhomogeneity in 2D
packings of geometrically confined particles, where Gaussian curvature is
induced and topological defects are excited to frustrate the crystalline
order~\cite{nelson2002defects,Mughal2007,yao2013topological,soni2018emergent}.
However, the widely used rotationally symmetric confining potentials in both
colloidal experiments and numerical simulations are incommensurate with the
triangular
lattice~\cite{yao2013topological,soni2018emergent,Klumov2019a,PhysRevE21Grason}.
This geometric incompatibility also leads to the frustration of the crystalline
order, and thus obscures the intrinsic role of the physical
interactions~\cite{thomas1994plasma,Leiderer1997,
Lai1999,Mughal2007,soni2018emergent}. It is therefore natural to inquire into
the assembly of particles purely dominated by the long-range interaction without
any complication from the boundary.  Revealing the intrinsic order in the
inhomogeneous particle packings yields insights into the long-range
interaction guided organization principle.

To address this question, we adopt the hexagonal geometry that is compatible
with the triangular lattice, and focus on the mechanical evolution of the
packing of the particles on a hexagonal disk under the long-range repulsive
force at zero temperature. Furthermore, we let the total number of
particles be a centered hexagonal number such that these particles may form a
defect-free triangular lattice within a hexagonal disk. As such, the hexagonal
disk model contains the essential elements for elucidating the intrinsic role of
the long-range interaction in organizing particles~\cite{yao2013topological}.

\begin{figure*}[th]  
\centering 
\includegraphics[width=6.8in]{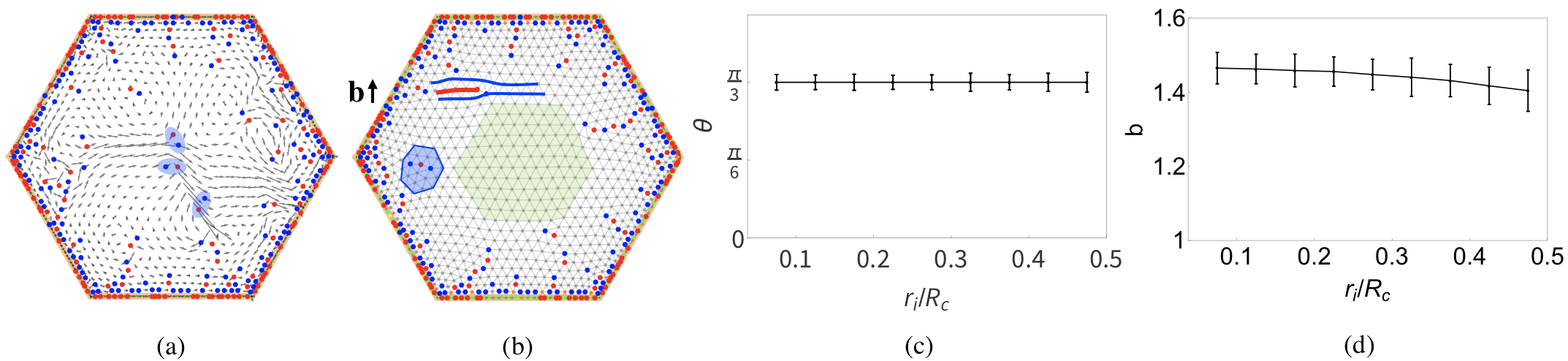}
  \caption{Emergence of the conformal defect-free zone (DFZ) in the
  inhomogeneous equilibrium packing of the particles confined on the hexagonal
  disk. (a) and (b) Full relaxation of the system from the intermediate state (a)
  to the equilibrium state (b). A characteristic event is the energetically
  favored vanishing of the three dislocations as highlighted by the ovals in (a).
  The DFZ is indicated by the light green hexagon in (b). The colored
  dots represent different kinds of disclinations; the red and blue dots in the
  interior of the disk refer to five- and seven-fold disclinations. Panels (c) and (d)
  show the constancy of the bond angle $\theta$ as a signature of conformal
  transformation, and the shrinking of the bond length $b$ towards the boundary.
  $R_c$ is the radius of the inscribed circle of the hexagonal boundary.
  $N=1141$. }
\label{conformal}
\end{figure*}

Based on the hexagonal disk model, we perform numerical experiment to fully
relax the particle configuration, and reveal the intrinsic conformal order in the
resulting inhomogeneous equilibrium packings of particles under both regular and
random initial conditions. Specifically, in the central defect-free zone (DFZ)
we observe the delicate angle-preserved bending of the lattice (without
breaking) to match the inhomogeneity condition.  Surrounding the central DFZ, we
discover the excitation of topological defects to resolve the intrinsic
geometric frustration. The revealed conformal structure as protected by the
surrounding defects represents a characteristic intrinsic order purely created
by the long-range repulsion.  These results advance our understanding on the
fundamental physics of long-range interacting systems, and may have practical
consequences in using long-range forces to control particle packings.

\section{Model and Method}

The model system consists of a collection of interacting particles confined on a
hexagonal disk. The interaction potential between particles is:
\begin{eqnarray}
  V(r) = \frac{\beta}{r^{\Gamma}},\label{Gamma}
\end{eqnarray}
where $\Gamma$ is an integer. The case of $\Gamma=1$ corresponds to the long-range Coulomb
potential in 3D space. We also investigate the cases of $\Gamma>1$ for shorter-range
interactions, and compare the results with the $1/r$-Coulomb interaction.
We shall point out that 2D Coulomb potential, as the Green's function
for the 2D Poisson equation, scales logarithmically with distance. Here, we
focus on the interaction potentials in the form of Eq.(\ref{Gamma}). One may
also resort to the screened Coulomb potential (also known as the Yukawa
potential) to investigate the effect of interaction range on the packing of
particles; the effective screened Coulomb potential originates from the ions in
electrolyte solution~\cite{yao2021epl}. Note that universal structural behavior
that is independent of the screening length has been reported in 2D Yukawa fluid
based on the analysis of the pair correlation function~\cite{Klumov2022b}.

Any particle that is outside the disk is subject to the confining potential:
\begin{eqnarray}
  V_{\textrm{conf.}}(r) = \frac{1}{2}k_0 d^2,
\end{eqnarray}
where $d$ is the shortest distance from the particle to the boundary, and
$k_0$ reflects the stiffness of the potential. In simulations, the value of
$k_0$ is set to be large enough [in comparison with $\beta$ in
Eq.~(\ref{Gamma})] to ensure that the particles are well confined
within the disk. Due to the independence of the lowest-energy configuration on the
value of $\beta$, we set $\beta=1$ in simulations.

We employ the standard steepest descent method to determine the
lowest-energy configurations in mechanical equilibrium without considering the
thermal effect.  The relaxation process consists of collective and individual
movements of the particles under the force on each particle. The typical step
size for these two kinds of relaxation modes are $s_c=10^{-4}a_0$ and
$s_i=0.1s_c$, respectively.  $a_0$ is the mean lattice spacing. Both regular and
random initial conditions are considered. In the regular initial state, the
particles in triangular lattice are arranged in concentric hexagons compatible
with the hexagonal boundary. The total number of particles is $N = 3n(n-1)+1$,
which is the $n$-th centered hexagonal number.  Detailed information about the
numerical simulation is provided in the Supplemental Material.

\section{RESULTS AND DISCUSSION}

\subsection{Identification of the central defect-free zone} 

We first discuss the
case of Coulomb potential with $\Gamma=1$ under the regular initial condition.
It is found that the initially perfect triangular lattice is disrupted under the
long-range repulsion as characterized by the excitation of the topological
defects [see Fig.~\ref{conformal}(a)]. The red and blue dots represent the
fundamental five- and seven-fold disclinations; the defects are identified by
the standard Delaunay triangulation procedure~\cite{nelson2002defects}. In a
triangular lattice, a $p$-fold disclination refers to a point in the
interior of the crystal with $p$ neighbors, and it carries a topological charge
of $q=6-z$~\cite{nelson2002defects}. A pair of five- and seven-fold
disclinations constitute a dislocation. The configuration in
Fig.~\ref{conformal}(a) is an intermediate state in the relaxation process. By
further relaxing the configuration via individually updating particle
positions, the system evolves towards the equilibrium state in
Fig.~\ref{conformal}(b). The displacement field is obtained from the deviation
of the configurations in Fig.~\ref{conformal}(a) and \ref{conformal}(b). The
displacement of the particles is represented by the arrows in
Figs.~\ref{conformal}(a), which constitute a ``flow" field. Comparison of
Figs.~\ref{conformal}(a) and \ref{conformal}(b) clearly shows that the central
defects [indicated by the ovals in Fig.~\ref{conformal}(a)] are swept away in
the ``flow".

A salient feature of the equilibrium configuration in Fig.~\ref{conformal}(b) is
the emergence of the defect-free zone (DFZ, as highlighted by the light green
hexagon) and the concentration of the geometric frustration near the boundary in
the form of topological defects, including dislocations and the 7-5-7 defect
strings that carry negative topological charge [as highlighted in the blue
hexagon]~\cite{nelson2002defects}. The defects are located where the lattice is
broken. For example, in Fig.~\ref{conformal}(b) the distance between the two
lattice lines (in blue) is enlarged by one lattice spacing as they meet a
dislocation (a pair of red and blue dots). The presence of the dislocation is to
introduce an extra line of particles (on the red line) between the two lattice
lines in blue.  Consequently, the contour integral of the
displacement field surrounding the dislocation in the lattice returns the
Burgers vector $\vec{b}$.  The direction of $\vec{b}$ is indicated by the black
arrow in Fig.~\ref{conformal}(b).

For the crystallized particle array confined on a circular disk, the geometric
incompatibility of the circular boundary and the interior triangular lattice
could lead to the emergence of
defects~\cite{Mughal2007,yao2013topological,soni2018emergent}. Here, in the
hexagonal system without any boundary caused geometric incompatibility, the
persistent boundary defects purely originate from the intrinsic density
inhomogeneity created by the long-range repulsion. In other words, the
increasing compression of the lattice towards the boundary, which is visible
in the triangulated particle configuration in Fig.~\ref{conformal}(b), leads to
the disruption of the boundary crystalline zone and the proliferation of the
defects. The long-range nature of the particle-particle interaction is crucial
for this process.

According to the connection between topological defects and stress, the
excitation of the boundary defects releases the stress in the
system~\cite{Chaikin1995}. Consequently, the crystalline order in the central
region is well preserved. In this sense, the central DFZ is protected by the
surrounding defects. Here, it is of interest to mention that in polydispersed
systems topological defects are excited surrounding an impurity particle of
larger size to protect the crystalline order in the exterior
region~\cite{yao2014polydispersity}.

Numerical experiments show that the emergence of the DFZ requires a full
relaxation of the system towards the energy valley with sufficiently fine step
size. The energy is reduced by as small as 0.00174\% as the system evolves from
Fig.~\ref{conformal}(a) to \ref{conformal}(b). The slight energy reduction is
consistent with the fact that the relaxation process is dominated by the glide
of dislocations, which requires a small amount of
energy~\cite{nelson2002defects}. Note that the packing of charged point
particles on a hexagonal disk has been investigated in our previous work,
focusing on the total amount of interior topological charge
$Q_{int}$~\cite{yao2013topological}. Here, we adopt the relaxation procedure
based on the combination of collective and individual movements of the particles
under a finer simulation step, and reveal the DFZ structure in the lowest-energy
particle configurations deeper in the energy valley. The total amount
of interior topological charge $Q_{int}$ is insensitive to the presence of the DFZ
in the center of the hexagonal disk system.

To check the reliability of the numerical result in Fig.~\ref{conformal}, we
further reduce the value for $s_i$ (the step size) down to $10^{-6}a_0$, and
obtain the identical defect pattern as in Fig.~\ref{conformal}(b). We
also adjust the boundary stiffness by tuning the value of $k_0$ in
Eq.(\ref{Gamma}), and find the persistence of the DFZ structure at varying
values of $k_0$.  The relevant lowest-energy particle configurations are
presented in the Supplemental Material.

\subsection{Conformality of the central defect-free zone}

To understand the formation of the DFZ in the environment of density
inhomogeneity, we first analyze the radial distributions of the bond angle and
the bond length. The results are presented in Figs.~\ref{conformal}(c) and
\ref{conformal}(d). It is found that from the center to the edge of the DFZ the
mean bond length is appreciably decreased by about $5\%$. In contrast, the mean
bond angle is a constant of $\pi/3$. As a signature of conformal transformation,
the preservation of the bond angle suggests that the triangular lattice in the
DFZ experiences a conformal deformation under the long-range repulsive force. 
Since the distribution of the bond angle is subject to a small fluctuation
around $\pi/3$, the DFZ is essentially a conformal crystal in the statistical
sense.

The notion of conformal crystal possesses an elegant mathematical structure
under the continuous medium approximation~\cite{pieranski1989gravity,rothen1993conformal,
rothen1996mechanical,wojciechowski1996minimum}. In general, the conformal
transformation is represented by an analytic function $\omega = f(z)$,
where $z=x+iy$. $\omega=u(x, y) + i v(x, y)$. Any point $(x_k, y_k)$ in the original
triangular lattice on the $(x, y)$-plane is mapped to the point $(u_k, v_k)$ on
the $(u, v)$-plane. The distance $ds_z$ of two points on the $(x, y)$-plane
becomes $ds_{\omega}$ on the $(u, v)$-plane. By some calculation, we have
\begin{eqnarray}
  ds_{\omega}^2= \left| \Lambda \right|^{2} ds_z^2,
  \label{ds2}
\end{eqnarray}
where $|\Lambda| = |d\omega/dz|$. By writing $ds_{\omega}^2 = ds_z^2 + 2 \epsilon_{\alpha\beta} dx_{\alpha}
dx_{\beta}$ where $x_1=x$ and $x_2=y$, and making use of the analyticity of $f(z)$, we obtain
\begin{eqnarray}
  \epsilon_{xx} &=& \epsilon_{yy},\nonumber \\ 
  \epsilon_{xy} &=& \epsilon_{yx} = 0. \label{epsilon}
\end{eqnarray}
The derivations of Eqs.(\ref{ds2}) and (\ref{epsilon}) are presented in the Supplemental Material.

Equation (\ref{epsilon}) indicates that the conformal crystal could be obtained
by a locally isotropic deformation (i.e., locally pure compression or stretching
without any shear) of a perfect triangular lattice. The information of the
density $n_{\omega}$ of the conformal crystal is encoded in the
$|\Lambda|$-factor. From the Jacobian of the transformation, we have 
\begin{eqnarray}
  n_{\omega} = \left| \Lambda \right|^{-2} n_z,\label{nn}
\end{eqnarray}
where $n_z$ is the density of the original triangular lattice. The analyticity
of the conformal transformation dictates that the logarithm of the density of
the conformal crystal conforms to the Laplace
equation~\cite{wojciechowski1996minimum}:
\begin{eqnarray} 
  \Delta \ln n_{\omega} = 0.\label{harmonic} 
\end{eqnarray}
Equations (\ref{epsilon}) and (\ref{harmonic}) imply that the formation of
conformal crystals requires subtle conditions. The deformation of the lattice
shall simultaneously satisfy the
density-inhomogeneity and the defect-free conditions. The concept of conformal
crystal has been introduced in a neat experiment, where a collection of
long-range repulsive magnetized spheres are self-organized into a regular
nonuniform distribution named gravity's rainbow~\cite{pieranski1989gravity,
rothen1996mechanical}. But strictly conformal crystals are difficult to be
realized; they could not be stabilized under uniform external
field~\cite{wojciechowski1996minimum}.

\begin{figure}[t]  
\centering 
\includegraphics[width=3.3in]{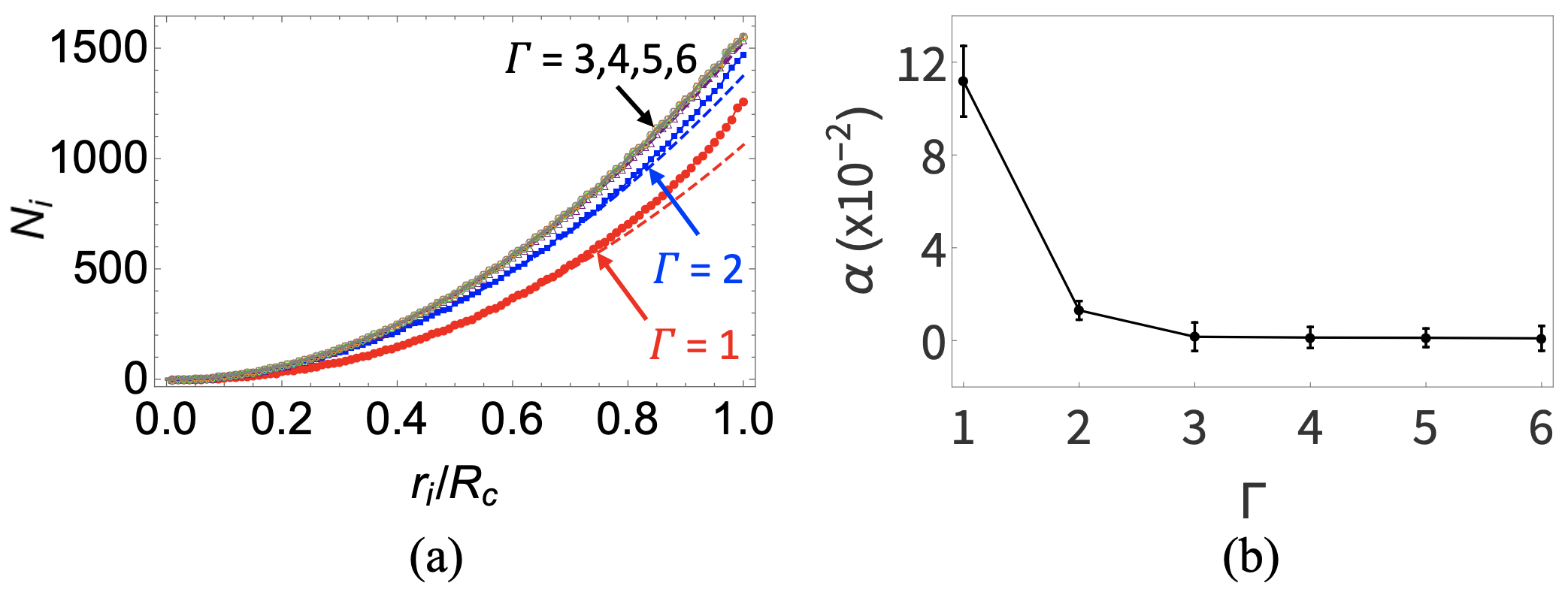}
  \caption{Conformality of the central defect-free zone as characterized by the
  power-law density distribution. (a) Cumulative particle distribution $N(r)$ in
  the equilibrium configurations for different values of $\Gamma$. $N(r) \propto
  r^{2+\alpha}$ in the DFZ. $\alpha=0.1$ and $0.013$ for $\Gamma=1$ and $2$,
  respectively. (b) The exponent $\alpha$ is plotted against the
  $\Gamma$-parameter. The error bars are obtained from the statistical analysis
  of the data for systems of varying $n$ ($n=20, 25, 30, 35, 40$). $N=1141$. }
\label{density}
\end{figure}

The rotationally symmetric solution to Eq.(\ref{harmonic}) conforms to:
$n_{\omega}(r) = C r^{\alpha}$, where $C$ and $\alpha$ are constants. As such,
in addition to the preserved bond angle, the power-law distribution of the
density in the deformed lattice is also a signature of conformal crystals. To
further examine the conformality of the DFZ, we analyze the radial distribution
of the confined particles, and present the main results in Fig.~\ref{density}.
$N_i$ is the number of particles within the circle of radius $r_i$. The particle
density $n_{\omega}(r)$ is related to the cumulative particle distribution
$N(r)$ by $n_{\omega}(r)=N'(r)/(2\pi r)$.

For the case of $\Gamma=1$, Fig.~\ref{density}(a) shows that the cumulative distribution
in the DFZ conforms to the power law: $N(r) \propto r^{2+\alpha}$, where
$\alpha=0.10$. Therefore, the logarithm of the particle density in the DFZ
conforms to the harmonic equation. The conformality of the DFZ is thus also
confirmed in terms of the density distribution. Note that, since $N(r) \propto
r^2$ for a uniform distribution, the parameter $\alpha$ serves as an indicator
of the degree of inhomogeneity. The power-law density profile can be realized by
the conformal transformation defined as $\omega = c'
z^{1-\alpha/2}$~\cite{silva2020formation}. The demonstration of the conformal
mappings is presented in the Supplemental Material.

  \begin{figure}[t]  
\centering 
\includegraphics[width=3in]{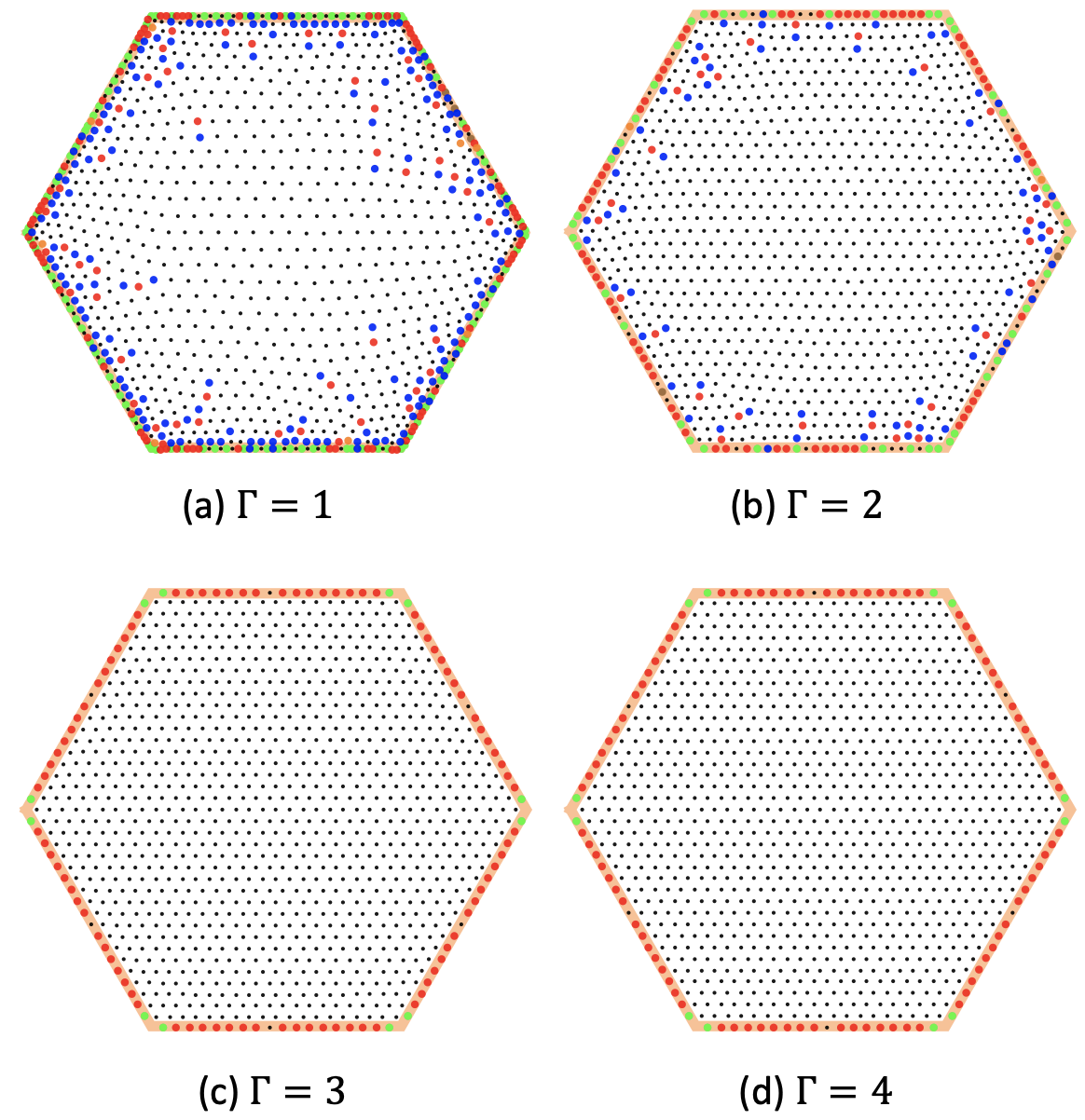}
  \caption{Lowest-energy particle configurations at varying $\Gamma$.
    Topological defects emerge for the cases $\Gamma=1$ and 2.  The red and blue
    dots in the interior of the disk represent five- and seven-fold
    disclinations. The equilibrium particle configurations for $\Gamma=3$ or
    larger are free of defects. $N=1141$. 
    }
\label{4Gamma}
\end{figure}

In Fig.~\ref{density}(a), we also present the cases of $\Gamma>1$. The
corresponding lowest-energy particle configurations at varying $\Gamma$ are
presented in Fig.~\ref{4Gamma}.  For $\Gamma=2$, similar to the case of
$\Gamma=1$, we observe the emergence of the DFZ that is surrounded by the
topological defects [see Fig.~\ref{4Gamma}(b)]. The corresponding cumulative
distribution in the DFZ could be well fitted by the power law with
$\alpha=0.013$. In contrast, Fig.~\ref{density}(a) shows that the data for the
cases of $\Gamma>2$ collapse on the same fitting curve of the quadratic
function. It indicates a homogeneous distribution of the particles, which is
confirmed in numerical experiment; the lowest-energy particle configurations
that are defect-free for $\Gamma=3$ and 4 are presented in Fig.~\ref{4Gamma}(c)
and \ref{4Gamma}(d). As such, the cases of $\Gamma \leq 2$ are fundamentally
different from the cases of larger $\Gamma$ in terms of homogeneity. In the
regime of long-range interaction ($\Gamma \leq 2$), the particle distribution
becomes inhomogeneous, producing richer structures like the DFZ and the boundary
defects.

\begin{figure*}[t]  
\centering 
\includegraphics[width=6.8in]{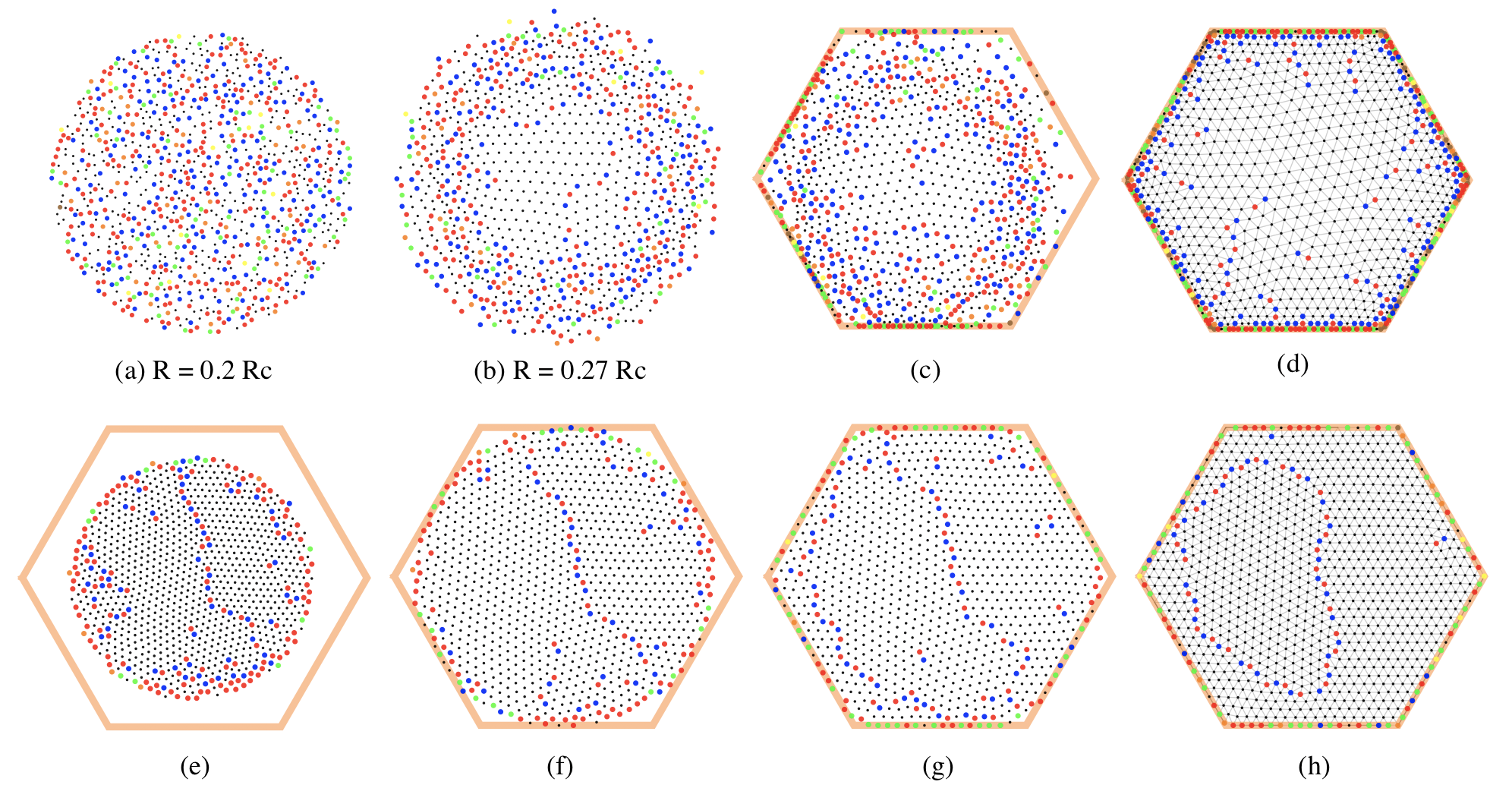}
  \caption{Ordering process in both long- and short-range interacting systems
  under the random initial condition. Panels (a)-(d) and (e)-(h) show typical
  snapshots in the relaxation of the particle configurations for $\Gamma=1$ and
  $6$, respectively. For both kinds of cases, the particles are initially
  randomly distributed within a circle of radius $R$, and the rightmost panels
  show the equilibrium configurations. In (d), we observe the emergence of the
  central defect-free zone in the long-range repulsive system under the random
  initial condition. In (h), dislocations are spontaneously aggregated into a
  loop structure in the regime of short-range repulsion.  $N=1000$. $R_c$ is the
  radius of the inscribed circle of the hexagonal boundary.
  }
\label{moving}
\end{figure*}

To investigate the dependence of the degree of inhomogeneity as indicated by the
value for $\alpha$ on the number of particles, we vary the value of $N$ in the
range from 1141 ($n=20$) to 4681 ($n=40$). The results are summarized 
in Fig.~\ref{density}(b).  The error bars are obtained from the
statistical analysis of the data for systems of varying $n$. We see that the
value for $\alpha$ is independent of the system size, and it is largely
determined by the range of physical interaction.

Now, we the analyze the characteristic distribution of density
in the DFZ from the geometric perspective. Equation~(\ref{ds2}) shows
that a line element in the original triangular lattice is isotropically expanded
by a factor of $|\Lambda|$ by the conformal transformation. The $|\Lambda|$-factor
thus contains the information about the Gaussian curvature of the resulting
conformal crystal according to the Gauss's Theorema Egregium. Given the metric
in Eq.~(\ref{ds2}), one can derive for
the Gaussian curvature~\cite{struik88a}: $K_G =
\frac{1}{|\Lambda|^2} \Delta \ln \left( \frac{1}{|\Lambda|}
  \right)$.
In terms of the
densities $n_{\omega}$ and $n_{z}$ by Eq.(\ref{nn}), we have
\begin{eqnarray}
  K_G = \frac{n_{\omega}}{2n_z} \Delta \ln \left( \frac{n_{\omega}}{n_z}
  \right).\label{KG}
\end{eqnarray} 
In combination with Eq.(\ref{harmonic}), we obtain an important
property of the conformal crystal that its Gaussian curvature is zero. In
general, non-uniform deformation of a perfect triangular lattice leads to the
variation of the metric, and thus gives rise to a nonzero intrinsic Gaussian
curvature. The conformal crystal sets the remarkable example of realizing zero
curvature in the environment of inhomogeneity.  Previous studies show that the
overall geometric effect of the long-range repulsion is to induce nonzero
Gaussian curvature in the inhomogeneous planar particle arrays~\cite{Nelson1987,
Mughal2007,yao2013topological,soni2018emergent}. Here, in the hexagonal system
we reveal the zero-curvature domain, where the bond-orientational order is well
preserved, and the particle distribution conforms to the power
law~\cite{nelson2002defects}.

We further analyze the DFZ structure by resorting to the fundamental relation
between particle density and topological charge density in two-dimensional
crystals~\cite{nelson2002defects}.  Geometric analysis shows that the
topological charge density is proportional to $\Delta \ln
n_{\omega}$~\cite{Mughal2007}. Therefore, Eq.(\ref{harmonic}) is also recognized
as the condition of zero topological charge density. To conclude, the preceding
discussions lead to a clear physical scenario about the organization of
geometrically confined long-range repulsive particles. In the central region,
the particles adopt the defect-free conformal organization, where the lattice
subtly bends to match the inhomogeneity condition without
breaking~\cite{Mughal2009}. The disruption of the crystalline order mainly
occurs near the boundary, where the excited topological defects serve as
stress-releasers to protect the conformal order in the central region.

Systematic simulations under a series of random initial conditions show that the
coexistence of the conformal DFZ and the defective boundary region is a common
phenomenon in the long-range interacting hexagonal systems. Specifically, we
confine randomly distributed particles within circular regions of varying radius
$R$. Let $R_c$ be the radius of the inscribed circle of the hexagonal boundary.
$R/R_c$ ranges from 0.2 to 0.9. Typical particle configurations in the
relaxation process are presented in the upper panels in Fig.~\ref{moving}.  A
key observation is that, in the free expansion of the circular cluster of
particles, ordering is initiated from the central region [see
Figs.~\ref{moving}(a) and \ref{moving}(b)]. The central region ultimately
becomes free of defects in the equilibrium state in Fig.~\ref{moving}(d). In all
of the cases with varying $R$, we uniformly observe the appearance of the DFZ in
the equilibrium configurations. In contrast, for short-range interacting
systems, by changing the initial condition from the regular triangular lattice
to random configurations, we observe the self-organization of dislocations into
the loop structure. Typical configurations in the relaxation process for
$\Gamma=6$ are presented in the lower panels in Fig.~\ref{moving}. The resulting
dislocation-fabricated loop separates the entire lattice into two crystallites
as shown in Fig.~\ref{moving}(h). More information about the formation of the
loop structure is provided in the Supplemental Material.

\section{Conclusion}

In summary, we investigate the fundamental question about the assembly of
particles under the long-range interaction, and reveal the intrinsic conformal
order in the inhomogeneous equilibrium packings. Specifically, the geometrically
confined particles are steered by the long-range repulsion to form a conformal
defect-free zone as protected by surrounding topological defects.  Around the
inhomogeneous defect-free zone, we discuss the remarkable confluence of the
deeply connected notions of conformality, harmonic inhomogeneity, zero
curvature, and zero topological charge density. This work demonstrates the
long-range interaction guided organization of matter, and may inspire further
explorations into the intriguing geometric structures underlying the
long-range interaction driven inhomogeneity.

\acknowledgments

This work was supported by the National Natural Science Foundation of
China (Grants No. BC4190050).


\begin{thebibliography}{40}
\expandafter\ifx\csname natexlab\endcsname\relax\def\natexlab#1{#1}\fi
\expandafter\ifx\csname bibnamefont\endcsname\relax
  \def\bibnamefont#1{#1}\fi
\expandafter\ifx\csname bibfnamefont\endcsname\relax
  \def\bibfnamefont#1{#1}\fi
\expandafter\ifx\csname citenamefont\endcsname\relax
  \def\citenamefont#1{#1}\fi
\expandafter\ifx\csname url\endcsname\relax
  \def\url#1{\texttt{#1}}\fi
\expandafter\ifx\csname urlprefix\endcsname\relax\def\urlprefix{URL }\fi
\providecommand{\bibinfo}[2]{#2}
\providecommand{\eprint}[2][]{\url{#2}}

\bibitem[{\citenamefont{Campa et~al.}(2014)\citenamefont{Campa, Dauxois,
  Fanelli, and Ruffo}}]{campa2014physics}
\bibinfo{author}{\bibfnamefont{A.}~\bibnamefont{Campa}},
  \bibinfo{author}{\bibfnamefont{T.}~\bibnamefont{Dauxois}},
  \bibinfo{author}{\bibfnamefont{D.}~\bibnamefont{Fanelli}}, \bibnamefont{and}
  \bibinfo{author}{\bibfnamefont{S.}~\bibnamefont{Ruffo}},
  \emph{\bibinfo{title}{Physics of Long-Range Interacting Systems}}
  (\bibinfo{publisher}{Oxford University Press, Oxford, UK},
  \bibinfo{year}{2014}).

\bibitem[{\citenamefont{Levin et~al.}(2014)\citenamefont{Levin, Pakter,
  Rizzato, Teles, and Benetti}}]{levin2014nonequilibrium}
\bibinfo{author}{\bibfnamefont{Y.}~\bibnamefont{Levin}},
  \bibinfo{author}{\bibfnamefont{R.}~\bibnamefont{Pakter}},
  \bibinfo{author}{\bibfnamefont{F.~B.} \bibnamefont{Rizzato}},
  \bibinfo{author}{\bibfnamefont{T.~N.} \bibnamefont{Teles}}, \bibnamefont{and}
  \bibinfo{author}{\bibfnamefont{F.~P.} \bibnamefont{Benetti}},
  \bibinfo{journal}{Phys. Rep.} \textbf{\bibinfo{volume}{535}},
  \bibinfo{pages}{1} (\bibinfo{year}{2014}).

\bibitem[{\citenamefont{Padmanabhan}(1990)}]{padmanabhan1990statistical}
\bibinfo{author}{\bibfnamefont{T.}~\bibnamefont{Padmanabhan}},
  \bibinfo{journal}{Phys. Rep.} \textbf{\bibinfo{volume}{188}},
  \bibinfo{pages}{285} (\bibinfo{year}{1990}).

\bibitem[{\citenamefont{Benetti et~al.}(2014)\citenamefont{Benetti,
  Ribeiro-Teixeira, Pakter, and Levin}}]{benetti2014nonequilibrium}
\bibinfo{author}{\bibfnamefont{F.~P.} \bibnamefont{Benetti}},
  \bibinfo{author}{\bibfnamefont{A.~C.} \bibnamefont{Ribeiro-Teixeira}},
  \bibinfo{author}{\bibfnamefont{R.}~\bibnamefont{Pakter}}, \bibnamefont{and}
  \bibinfo{author}{\bibfnamefont{Y.}~\bibnamefont{Levin}},
  \bibinfo{journal}{Phys. Rev. Lett.} \textbf{\bibinfo{volume}{113}},
  \bibinfo{pages}{100602} (\bibinfo{year}{2014}).

\bibitem[{\citenamefont{Thomas et~al.}(1994)\citenamefont{Thomas, Morfill,
  Demmel, Goree, Feuerbacher, and M{\"o}hlmann}}]{thomas1994plasma}
\bibinfo{author}{\bibfnamefont{H.}~\bibnamefont{Thomas}},
  \bibinfo{author}{\bibfnamefont{G.}~\bibnamefont{Morfill}},
  \bibinfo{author}{\bibfnamefont{V.}~\bibnamefont{Demmel}},
  \bibinfo{author}{\bibfnamefont{J.}~\bibnamefont{Goree}},
  \bibinfo{author}{\bibfnamefont{B.}~\bibnamefont{Feuerbacher}},
  \bibnamefont{and}
  \bibinfo{author}{\bibfnamefont{D.}~\bibnamefont{M{\"o}hlmann}},
  \bibinfo{journal}{Phys. Rev. Lett.} \textbf{\bibinfo{volume}{73}},
  \bibinfo{pages}{652} (\bibinfo{year}{1994}).

\bibitem[{\citenamefont{Levin}(2002)}]{Levin2002}
\bibinfo{author}{\bibfnamefont{Y.}~\bibnamefont{Levin}}, \bibinfo{journal}{Rep.
  Prog. Phys.} \textbf{\bibinfo{volume}{65}}, \bibinfo{pages}{1577}
  (\bibinfo{year}{2002}).

\bibitem[{\citenamefont{Landau and Lifshitz}(1986)}]{Landau1986}
\bibinfo{author}{\bibfnamefont{L.~D.} \bibnamefont{Landau}} \bibnamefont{and}
  \bibinfo{author}{\bibfnamefont{E.~M.} \bibnamefont{Lifshitz}},
  \emph{\bibinfo{title}{Theory of Elasticity, 3rd edition}}
  (\bibinfo{publisher}{Butterworth-Heinemann}, \bibinfo{year}{1986}).

\bibitem[{\citenamefont{Audoly and Pomeau}(2010)}]{audoly2010elasticity}
\bibinfo{author}{\bibfnamefont{B.}~\bibnamefont{Audoly}} \bibnamefont{and}
  \bibinfo{author}{\bibfnamefont{Y.}~\bibnamefont{Pomeau}},
  \emph{\bibinfo{title}{Elasticity and geometry}} (\bibinfo{publisher}{Oxford
  Univ. Press}, \bibinfo{year}{2010}).

\bibitem[{\citenamefont{Chattopadhyay and Wu}(2009)}]{chattopadhyay2009effect}
\bibinfo{author}{\bibfnamefont{S.}~\bibnamefont{Chattopadhyay}}
  \bibnamefont{and} \bibinfo{author}{\bibfnamefont{X.-L.} \bibnamefont{Wu}},
  \bibinfo{journal}{Biophys. J.} \textbf{\bibinfo{volume}{96}},
  \bibinfo{pages}{2023} (\bibinfo{year}{2009}).

\bibitem[{\citenamefont{Dallaston et~al.}(2018)\citenamefont{Dallaston,
  Fontelos, Tseluiko, and Kalliadasis}}]{dallaston2018discrete}
\bibinfo{author}{\bibfnamefont{M.~C.} \bibnamefont{Dallaston}},
  \bibinfo{author}{\bibfnamefont{M.~A.} \bibnamefont{Fontelos}},
  \bibinfo{author}{\bibfnamefont{D.}~\bibnamefont{Tseluiko}}, \bibnamefont{and}
  \bibinfo{author}{\bibfnamefont{S.}~\bibnamefont{Kalliadasis}},
  \bibinfo{journal}{Phys. Rev. Lett.} \textbf{\bibinfo{volume}{120}},
  \bibinfo{pages}{034505} (\bibinfo{year}{2018}).

\bibitem[{\citenamefont{Holm et~al.}(2001)\citenamefont{Holm, K{\'e}kicheff,
  and Podgornik}}]{Holm2001}
\bibinfo{author}{\bibfnamefont{C.}~\bibnamefont{Holm}},
  \bibinfo{author}{\bibfnamefont{P.}~\bibnamefont{K{\'e}kicheff}},
  \bibnamefont{and}
  \bibinfo{author}{\bibfnamefont{R.}~\bibnamefont{Podgornik}},
  \emph{\bibinfo{title}{Electrostatic Effects in Soft Matter and Biophysics}}
  (\bibinfo{publisher}{Springer, Berlin}, \bibinfo{year}{2001}).

\bibitem[{\citenamefont{Walker et~al.}(2011)\citenamefont{Walker, Kowalczyk,
  Olvera de~la Cruz, and Grzybowski}}]{Walker2011}
\bibinfo{author}{\bibfnamefont{D.~A.} \bibnamefont{Walker}},
  \bibinfo{author}{\bibfnamefont{B.}~\bibnamefont{Kowalczyk}},
  \bibinfo{author}{\bibfnamefont{M.}~\bibnamefont{Olvera de~la Cruz}},
  \bibnamefont{and} \bibinfo{author}{\bibfnamefont{B.~A.}
  \bibnamefont{Grzybowski}}, \bibinfo{journal}{Nanoscale}
  \textbf{\bibinfo{volume}{3}}, \bibinfo{pages}{1316} (\bibinfo{year}{2011}).

\bibitem[{\citenamefont{Lee et~al.}(2015)\citenamefont{Lee, Waitukaitis,
  Miskin, and Jaeger}}]{lee2015direct}
\bibinfo{author}{\bibfnamefont{V.}~\bibnamefont{Lee}},
  \bibinfo{author}{\bibfnamefont{S.~R.} \bibnamefont{Waitukaitis}},
  \bibinfo{author}{\bibfnamefont{M.~Z.} \bibnamefont{Miskin}},
  \bibnamefont{and} \bibinfo{author}{\bibfnamefont{H.~M.}
  \bibnamefont{Jaeger}}, \bibinfo{journal}{Nat. Phys.}
  \textbf{\bibinfo{volume}{11}}, \bibinfo{pages}{733} (\bibinfo{year}{2015}).

\bibitem[{\citenamefont{Shen et~al.}(2019)\citenamefont{Shen, Tong, Tan, and
  Xu}}]{shen2019universal}
\bibinfo{author}{\bibfnamefont{H.}~\bibnamefont{Shen}},
  \bibinfo{author}{\bibfnamefont{H.}~\bibnamefont{Tong}},
  \bibinfo{author}{\bibfnamefont{P.}~\bibnamefont{Tan}}, \bibnamefont{and}
  \bibinfo{author}{\bibfnamefont{L.}~\bibnamefont{Xu}}, \bibinfo{journal}{Nat.
  Commun.} \textbf{\bibinfo{volume}{10}}, \bibinfo{pages}{1}
  (\bibinfo{year}{2019}).

\bibitem[{\citenamefont{Schweigert and Peeters}(1995)}]{schweigert1995spectral}
\bibinfo{author}{\bibfnamefont{V.~A.} \bibnamefont{Schweigert}}
  \bibnamefont{and} \bibinfo{author}{\bibfnamefont{F.~M.}
  \bibnamefont{Peeters}}, \bibinfo{journal}{Phys. Rev. B}
  \textbf{\bibinfo{volume}{51}}, \bibinfo{pages}{7700} (\bibinfo{year}{1995}).

\bibitem[{\citenamefont{O{\u{g}}uz et~al.}(2011)\citenamefont{O{\u{g}}uz,
  Messina, and L{\"o}wen}}]{ouguz2011helicity}
\bibinfo{author}{\bibfnamefont{E.}~\bibnamefont{O{\u{g}}uz}},
  \bibinfo{author}{\bibfnamefont{R.}~\bibnamefont{Messina}}, \bibnamefont{and}
  \bibinfo{author}{\bibfnamefont{H.}~\bibnamefont{L{\"o}wen}},
  \bibinfo{journal}{Europhys. Lett.} \textbf{\bibinfo{volume}{94}},
  \bibinfo{pages}{28005} (\bibinfo{year}{2011}).

\bibitem[{\citenamefont{Ribeiro-Teixeira
  et~al.}(2014)\citenamefont{Ribeiro-Teixeira, Benetti, Pakter, and
  Levin}}]{ribeiro2014ergodicity}
\bibinfo{author}{\bibfnamefont{A.~C.} \bibnamefont{Ribeiro-Teixeira}},
  \bibinfo{author}{\bibfnamefont{F.~P.} \bibnamefont{Benetti}},
  \bibinfo{author}{\bibfnamefont{R.}~\bibnamefont{Pakter}}, \bibnamefont{and}
  \bibinfo{author}{\bibfnamefont{Y.}~\bibnamefont{Levin}},
  \bibinfo{journal}{Phys. Rev. E} \textbf{\bibinfo{volume}{89}},
  \bibinfo{pages}{022130} (\bibinfo{year}{2014}).

\bibitem[{\citenamefont{Berezin}(1985)}]{Berezin1985}
\bibinfo{author}{\bibfnamefont{A.}~\bibnamefont{Berezin}},
  \bibinfo{journal}{Nature} \textbf{\bibinfo{volume}{315}},
  \bibinfo{pages}{104} (\bibinfo{year}{1985}).

\bibitem[{\citenamefont{Yao and Olvera de~la Cruz}(2013)}]{yao2013topological}
\bibinfo{author}{\bibfnamefont{Z.}~\bibnamefont{Yao}} \bibnamefont{and}
  \bibinfo{author}{\bibfnamefont{M.}~\bibnamefont{Olvera de~la Cruz}},
  \bibinfo{journal}{Phys. Rev. Lett.} \textbf{\bibinfo{volume}{111}},
  \bibinfo{pages}{115503} (\bibinfo{year}{2013}).

\bibitem[{\citenamefont{Soni et~al.}(2018)\citenamefont{Soni, G{\'o}mez, and
  Irvine}}]{soni2018emergent}
\bibinfo{author}{\bibfnamefont{V.}~\bibnamefont{Soni}},
  \bibinfo{author}{\bibfnamefont{L.~R.} \bibnamefont{G{\'o}mez}},
  \bibnamefont{and} \bibinfo{author}{\bibfnamefont{W.~T.}
  \bibnamefont{Irvine}}, \bibinfo{journal}{Phys. Rev. X}
  \textbf{\bibinfo{volume}{8}}, \bibinfo{pages}{011039} (\bibinfo{year}{2018}).

\bibitem[{\citenamefont{Silva et~al.}(2020)\citenamefont{Silva, Menezes,
  Cabral, and de~Souza~Silva}}]{silva2020formation}
\bibinfo{author}{\bibfnamefont{F.~C.} \bibnamefont{Silva}},
  \bibinfo{author}{\bibfnamefont{R.~M.} \bibnamefont{Menezes}},
  \bibinfo{author}{\bibfnamefont{L.~R.} \bibnamefont{Cabral}},
  \bibnamefont{and} \bibinfo{author}{\bibfnamefont{C.~C.}
  \bibnamefont{de~Souza~Silva}}, \bibinfo{journal}{J. Phys.: Condens. Matter}
  \textbf{\bibinfo{volume}{32}}, \bibinfo{pages}{505401}
  (\bibinfo{year}{2020}).

\bibitem[{\citenamefont{Sun and Yao}(2023)}]{PhysRevE.108.025001}
\bibinfo{author}{\bibfnamefont{H.}~\bibnamefont{Sun}} \bibnamefont{and}
  \bibinfo{author}{\bibfnamefont{Z.}~\bibnamefont{Yao}},
  \bibinfo{journal}{Phys. Rev. E} \textbf{\bibinfo{volume}{108}},
  \bibinfo{pages}{025001} (\bibinfo{year}{2023}).

\bibitem[{\citenamefont{Mughal and Weaire}(2009)}]{Mughal2009}
\bibinfo{author}{\bibfnamefont{A.}~\bibnamefont{Mughal}} \bibnamefont{and}
  \bibinfo{author}{\bibfnamefont{D.}~\bibnamefont{Weaire}},
  \bibinfo{journal}{Proc. R. Soc. London, Ser. A}
  \textbf{\bibinfo{volume}{465}}, \bibinfo{pages}{219} (\bibinfo{year}{2009}).

\bibitem[{\citenamefont{Lai and Lin}(1999)}]{Lai1999}
\bibinfo{author}{\bibfnamefont{Y.-J.} \bibnamefont{Lai}} \bibnamefont{and}
  \bibinfo{author}{\bibfnamefont{I.}~\bibnamefont{Lin}},
  \bibinfo{journal}{Phys. Rev. E} \textbf{\bibinfo{volume}{60}},
  \bibinfo{pages}{4743} (\bibinfo{year}{1999}).

\bibitem[{\citenamefont{Nelson}(2002)}]{nelson2002defects}
\bibinfo{author}{\bibfnamefont{D.~R.} \bibnamefont{Nelson}},
  \emph{\bibinfo{title}{Defects and Geometry in Condensed Matter Physics}}
  (\bibinfo{publisher}{Cambridge University Press, Cambridge},
  \bibinfo{year}{2002}).

\bibitem[{\citenamefont{Mughal and Moore}(2007)}]{Mughal2007}
\bibinfo{author}{\bibfnamefont{A.}~\bibnamefont{Mughal}} \bibnamefont{and}
  \bibinfo{author}{\bibfnamefont{M.}~\bibnamefont{Moore}},
  \bibinfo{journal}{Phys. Rev. E} \textbf{\bibinfo{volume}{76}},
  \bibinfo{pages}{011606} (\bibinfo{year}{2007}).

\bibitem[{\citenamefont{Klumov}(2022{\natexlab{a}})}]{Klumov2022}
\bibinfo{author}{\bibfnamefont{B.~A.} \bibnamefont{Klumov}},
  \bibinfo{journal}{JETP Letters} \textbf{\bibinfo{volume}{116}},
  \bibinfo{pages}{703} (\bibinfo{year}{2022}{\natexlab{a}}).

\bibitem[{\citenamefont{Piera{\'n}ski}(1989)}]{pieranski1989gravity}
\bibinfo{author}{\bibfnamefont{P.}~\bibnamefont{Piera{\'n}ski}}, in
  \emph{\bibinfo{booktitle}{Phase Transitions in Soft Condensed Matter}}
  (\bibinfo{publisher}{Springer}, \bibinfo{year}{1989}), pp.
  \bibinfo{pages}{45--48}.

\bibitem[{\citenamefont{Rothen et~al.}(1993)\citenamefont{Rothen, Pieranski,
  Rivier, and Joyet}}]{rothen1993conformal}
\bibinfo{author}{\bibfnamefont{F.}~\bibnamefont{Rothen}},
  \bibinfo{author}{\bibfnamefont{P.}~\bibnamefont{Pieranski}},
  \bibinfo{author}{\bibfnamefont{N.}~\bibnamefont{Rivier}}, \bibnamefont{and}
  \bibinfo{author}{\bibfnamefont{A.}~\bibnamefont{Joyet}},
  \bibinfo{journal}{Eur. J. Phys.} \textbf{\bibinfo{volume}{14}},
  \bibinfo{pages}{227} (\bibinfo{year}{1993}).

\bibitem[{\citenamefont{Rothen and Piera{\'n}ski}(1996)}]{rothen1996mechanical}
\bibinfo{author}{\bibfnamefont{F.}~\bibnamefont{Rothen}} \bibnamefont{and}
  \bibinfo{author}{\bibfnamefont{P.}~\bibnamefont{Piera{\'n}ski}},
  \bibinfo{journal}{Phys. Rev. E} \textbf{\bibinfo{volume}{53}},
  \bibinfo{pages}{2828} (\bibinfo{year}{1996}).

\bibitem[{\citenamefont{Wojciechowski and
  Klos}(1996)}]{wojciechowski1996minimum}
\bibinfo{author}{\bibfnamefont{K.~W.} \bibnamefont{Wojciechowski}}
  \bibnamefont{and} \bibinfo{author}{\bibfnamefont{J.}~\bibnamefont{Klos}},
  \bibinfo{journal}{J. Phys. A: Math. Gen.} \textbf{\bibinfo{volume}{29}},
  \bibinfo{pages}{3963} (\bibinfo{year}{1996}).

\bibitem[{\citenamefont{Klumov}(2019)}]{Klumov2019a}
\bibinfo{author}{\bibfnamefont{B.~A.} \bibnamefont{Klumov}},
  \bibinfo{journal}{JETP Letters} \textbf{\bibinfo{volume}{110}},
  \bibinfo{pages}{715} (\bibinfo{year}{2019}).

\bibitem[{\citenamefont{Meng and Grason}(2021)}]{PhysRevE21Grason}
\bibinfo{author}{\bibfnamefont{Q.}~\bibnamefont{Meng}} \bibnamefont{and}
  \bibinfo{author}{\bibfnamefont{G.~M.} \bibnamefont{Grason}},
  \bibinfo{journal}{Phys. Rev. E} \textbf{\bibinfo{volume}{104}},
  \bibinfo{pages}{034614} (\bibinfo{year}{2021}).

\bibitem[{\citenamefont{S.~Neser and Leiderer}(1997)}]{Leiderer1997}
\bibinfo{author}{\bibfnamefont{C.~B.} \bibnamefont{S.~Neser},
  \bibfnamefont{T.~Palberg}} \bibnamefont{and}
  \bibinfo{author}{\bibfnamefont{P.}~\bibnamefont{Leiderer}},
  \bibinfo{journal}{Prog. Colloid Polym. Sci.} \textbf{\bibinfo{volume}{104}},
  \bibinfo{pages}{194} (\bibinfo{year}{1997}).

\bibitem[{\citenamefont{Yao}(2021)}]{yao2021epl}
\bibinfo{author}{\bibfnamefont{Z.}~\bibnamefont{Yao}},
  \bibinfo{journal}{Europhys. Lett.} \textbf{\bibinfo{volume}{133}},
  \bibinfo{pages}{54002} (\bibinfo{year}{2021}).

\bibitem[{\citenamefont{Klumov}(2022{\natexlab{b}})}]{Klumov2022b}
\bibinfo{author}{\bibfnamefont{B.~A.} \bibnamefont{Klumov}},
  \bibinfo{journal}{JETP Letters} \textbf{\bibinfo{volume}{115}},
  \bibinfo{pages}{108} (\bibinfo{year}{2022}{\natexlab{b}}).

\bibitem[{\citenamefont{Chaikin and Lubensky}(1995)}]{Chaikin1995}
\bibinfo{author}{\bibfnamefont{P.}~\bibnamefont{Chaikin}} \bibnamefont{and}
  \bibinfo{author}{\bibfnamefont{T.}~\bibnamefont{Lubensky}},
  \emph{\bibinfo{title}{Principles of Condensed Matter Physics}}
  (\bibinfo{publisher}{Cambridge University Press}, \bibinfo{year}{1995}).

\bibitem[{\citenamefont{Yao and Olvera de~la
  Cruz}(2014)}]{yao2014polydispersity}
\bibinfo{author}{\bibfnamefont{Z.}~\bibnamefont{Yao}} \bibnamefont{and}
  \bibinfo{author}{\bibfnamefont{M.}~\bibnamefont{Olvera de~la Cruz}},
  \bibinfo{journal}{Proc. Natl. Acad. Sci. U.S.A.}
  \textbf{\bibinfo{volume}{111}}, \bibinfo{pages}{5094} (\bibinfo{year}{2014}).

\bibitem[{\citenamefont{Struik}(1988)}]{struik88a}
\bibinfo{author}{\bibfnamefont{D.}~\bibnamefont{Struik}},
  \emph{\bibinfo{title}{Lectures on Classical Differential Geometry}}
  (\bibinfo{publisher}{Dover Publications, New York}, \bibinfo{year}{1988}),
  \bibinfo{edition}{2nd} ed.

\bibitem[{\citenamefont{Nelson and Peliti}(1987)}]{Nelson1987}
\bibinfo{author}{\bibfnamefont{D.}~\bibnamefont{Nelson}} \bibnamefont{and}
  \bibinfo{author}{\bibfnamefont{L.}~\bibnamefont{Peliti}},
  \bibinfo{journal}{J. Phys. (France)} \textbf{\bibinfo{volume}{48}},
  \bibinfo{pages}{1085} (\bibinfo{year}{1987}).

\end{thebibliography}

\end{document}